\documentclass[
12pt,%
a4paper,
notitlepage,
oneside,%
onecolumn]{article}
\usepackage{graphicx}
\usepackage{color}

\begin{document}
\title{On the contribution of the $^{40}$K geo-antineutrino to single Borexino events. \\ Version 2}
\author{ L.B. Bezrukov, I.S. Karpikov, A.S. Kurlovich, A.K. Mezhokh,\\
 S.V.Silaeva, V.V. Sinev and V.P. Zavarzina }
\maketitle
 Institute for Nuclear Research of Russian academy of sciences, Moscow

\begin{abstract} 
We propose to include in the analysis of Borexino single event energy spectrum the scattering of $^{40}$K geo-antineutrinos by scintillator electrons. The Hydridic Earth model predicts the concentration of potassium in modern Earth from 1\% to 4\% of the Earth mass. We calculated contribution of $^{40}$K geo-antineutrino interactions in single Borexino events for these concentrations. This contribution is comparable to the contribution from the interaction of CNO neutrinos.  We discuss the reasons for using the Hydridic Earth model.                   
\end{abstract} 

\section{Introduction.}
\hspace{0.5cm}
Recently the Borexino collaboration \cite{agost} published the results of analysis of the experimental energy spectrum of single events. In this work the authors fitted the experimental spectrum with the theoretical one.  As a result, the interaction rates from various sources were obtained except for CNO solar neutrinos interaction rate. 

To obtain the values of the interaction rates from all sources, it is necessary to have the correct theoretical information about the shape of the measured energy spectrum from each of the possible sources. In the absence of correct information, the used procedure for fitting of experimental data with the model may not allow to obtain the interaction rates and their uncertainties from the sources used in the model. 

The results of \cite{agost} most likely demonstrate the lack of complete information on possible sources of single events. The paper reports that there is no way to get the interaction rate of CNO neutrinos. Only the upper bound of $8\div10$ events per day in 100 tons of scintillator was obtained.  From 3 to 5 events per day are expected from CNO neutrino interactions depending on the model of the Sun.  

\section{Energy spectrum of recoil electrons from scattering of $^{40}$K geo-antineutrino.}
\hspace{0.5cm} 
The authors of the work \cite{bezr} considered an additional source of single events for the Borexino detector. This is the scattering of potassium geo-antineutrinos by electrons. Such a source was not considered in \cite{agost} because the Silicate Earth model was used in it. The concentration of potassium in this model is very small (0.024\% of the Earth mass) and the contribution to single events of the Borexino detector from potassium geo-antineutrinos is negligible.
       
\includegraphics[scale=0.7]{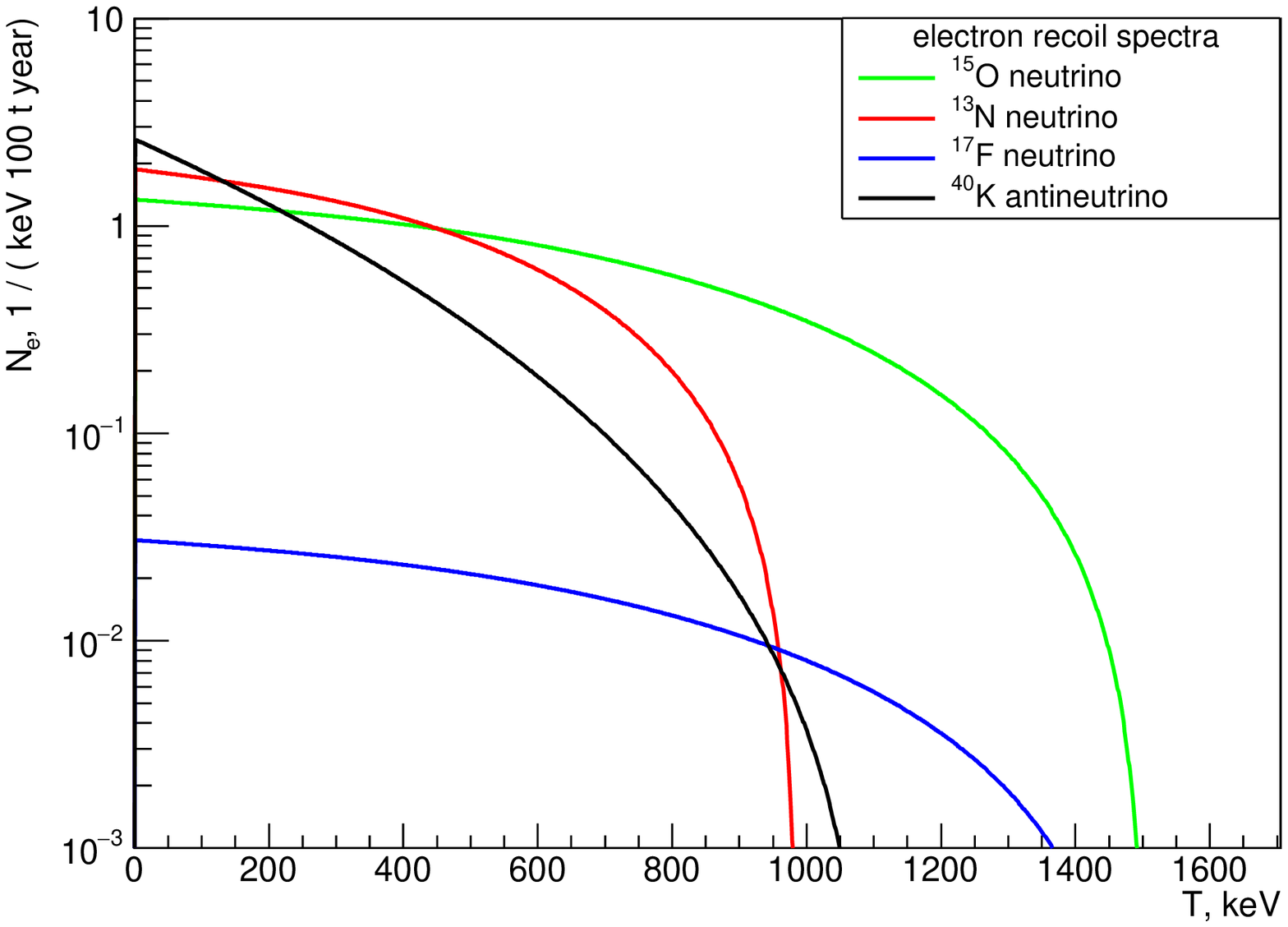}

\small{Fig.1. Differential energy spectra of recoil electrons from neutrino  (or antineutrino) scattering on electrons for various sources (the oscillations were not taken into account). The red line is the solar neutrino from $^{13}$N decay in CNO cycle. The green is the solar neutrino from $^{15}$O decay in CNO cycle. The blue is the solar neutrino from $^{17}$F decay in CNO cycle. The black is $^{40}$K geo-antineutrino for the concentration of potassium in modern Earth 1\% of the Earth mass}

\includegraphics[scale=0.7]{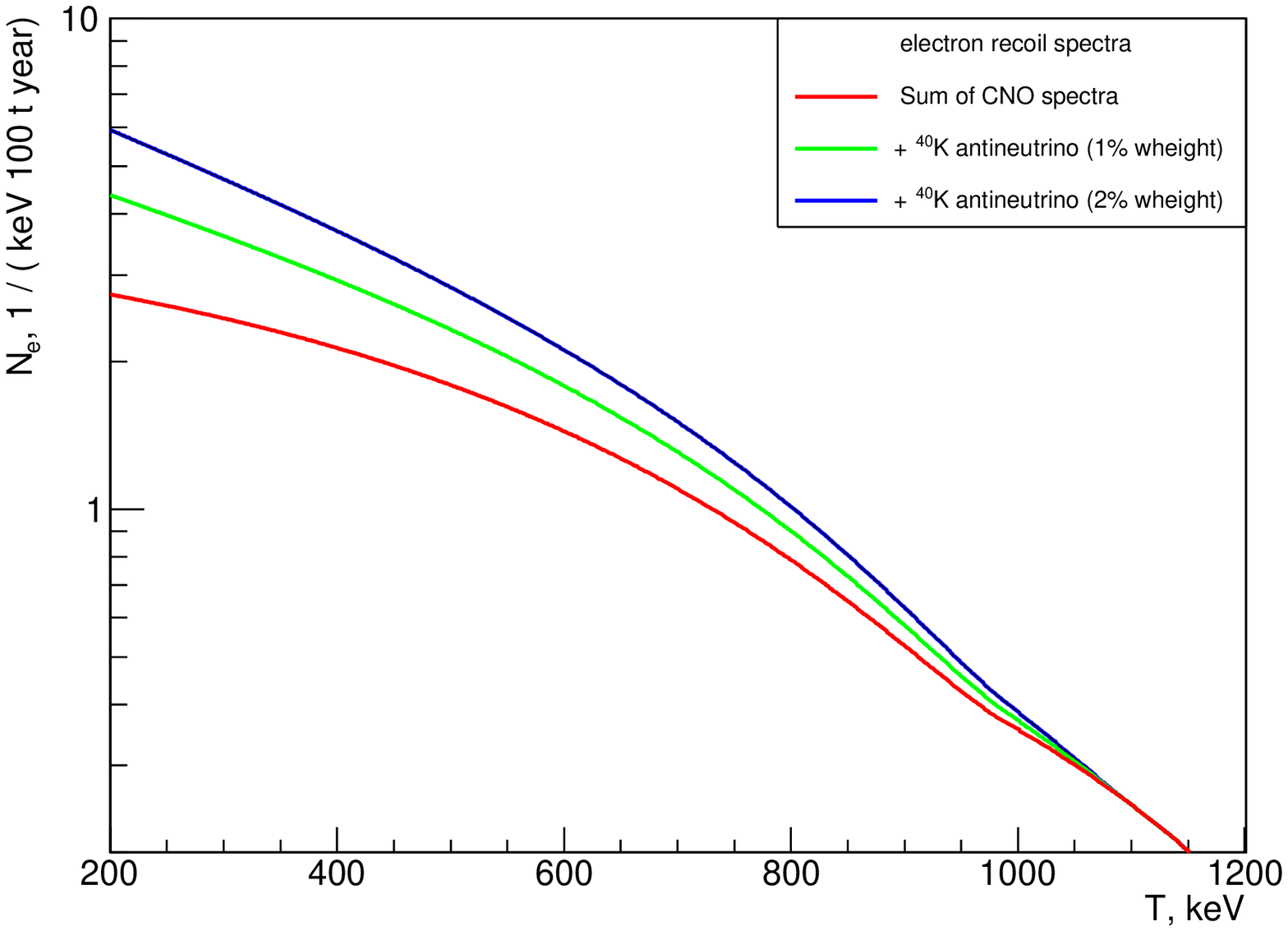}

\small{Fig.2. Differential energy spectra of recoil electrons from solar CNO neutrino scattering by scintillator electrons in 100 tons of scintillator (red curve) and the sum of the recoil electron spectra from the scattering of solar CNO neutrino  and from the scattering of  $^{40}$K geo-antineutrino for the concentration of potassium in the modern Earth from 1\% to 2\% of the Earth mass. The oscillations were taken into account.}
\vspace{0.5cm}

\hspace{0.5cm} 
We proceeded from the prediction of the Earth elemental composition by another model of the Earth "Hydridic Earth" or "Hydrogen-rich Earth" \cite{Larin, toul}. These studies predict concentration of potassium in the original Earth up to 3.8\% of the Earth mass. In the modern Earth we can expect a larger proportion of potassium due to the loss of hydrogen by the Earth during the time of it's existence. The predictions of the Hydridic Earth model have low accuracy, so it is reasonable to search for the flux of potassium geo-antineutrinos in a wide range of values.  

The work \cite{bezr}  presented the results of calculation of the energy spectrum of recoil electrons from geo-antineutrino scattering from $^{40}$K decay on scintillator electrons and the counting rate of these events in 100 tons of scintillator for the concentration of potassium in the modern Earth from 0.3\% to 3.8\% of the Earth's mass. 

We present at figure 1 our calculation (using the approximations described in  \cite{bezr} and the oscillations were not taken into account) of differential energy spectra of recoil electrons produced by the scattering of neutrinos  (or antineutrinos) on electrons for different sources: red curve - solar neutrinos from the decay of $^{13}$N in the CNO cycle, the green curve corresponds to the solar neutrinos from the decay of $^{15}$O in the CNO cycle, blue curve - solar neutrinos from the decay of  $^{17}$F in the CNO cycle, black curve -  geo-antineutrinos from $^{40}$K decay for the concentration of potassium in the Earth 1\% of weight of the Earth. We can see that the spectrum of $^{40}$K geo-antineutrinos for the concentration of potassium in the Earth of 1\% of the Earth mass is similar to the spectrum of solar neutrinos from the decay of $^{13}$N in the CNO cycle. It has an upper limit larger than that of the spectrum of solar neutrinos from the decay of $^{13}$N. For our calculation we used spectra of solar neutrinos from the CNO cycle \cite{Bahcall} — original work of J. Bahcall, which are normalized to the value of the flux for large metallicity of the Sun ($4.88\cdot10^8$ cm$^{-2}$ s$^{-1}$).

Figure 2 shows the differential energy spectrum from CNO neutrino scattering by electrons in 100 tons of scintillator (red line) and the total recoil electron spectrum from the scattering of CNO neutrino and $^{40}$K geo-antineutrino for the concentration of potassium in the modern Earth from 1\% to 2\% of the Earth mass (green and blue curves respectively). The oscillations were taken into account. It means that the contribution of muon and tau neutrinos (and anty-neutrinos) scattering was taken into account. 

 The intensity of events from the CNO neutrino calculated for the curve from figure 2 is 4.9 cpd/100 tons, and the additional intensities of events from $^{40}$K geo-antineutrinos is 1.6 to 3.4 cpd/100 tons.
The Borexino collaboration in \cite{agost} used a differential energy spectrum of recoil electrons from the scattering of  CNO  neutrinos on electrons, similar to the one shown in figure 2 in red, and did not obtain expected value of interaction rate ($3 \div 5$ cpd/100 tons). 

We propose to perform the same study as in Borexino collaboration work \cite{agost} using the total spectrum of recoil electrons from the scattering of CNO neutrinos and from the scattering of $^{40}$K geo-antineutrino for the potassium concentration in the modern Earth from 1\% to 2\% of the Earth's mass. We expect to find a certain value for the intensity of total events from the CNO neutrinos and from the $^{40}$K geo-antineutrinos in the range of $6 \div 9$ events per day in 100 tons of scintillator. We also expect that, as a result of this analysis, several events per day will be taken away from events attributed in work  \cite{agost} to solar $^{7}$Be neutrinos interactions and $^{210}$Bi decays. The errors of the obtained intensities will become symmetrical.

We give the functions from figure 2 in table form (see table below).

The obtaining the intensity of total spectrum of recoil electrons from the scattering of CNO neutrinos and $^{40}$K geo-antineutrinos as a result of fitting analysis of Borexino single event data will be an important confirmation of the validity of the Hydridic Earth model. 

\begin{table}[ht]
\caption{Differential energy spectra of recoil electrons from solar CNO neutrino scattering by scintillator electrons in 100 tons of scintillator and the same with addition of $^{40}$K antineutrinos in units $MeV^{-1} year^{-1} (100 tons)^{-1}$. }
\label{tabl:event}
\centering
\vspace{2mm}
 \begin{tabular}{| c | c | c | c | c |} 
 \hline
 T, MeV &  CNO  & CNO+1\%($^{40}$K)  & CNO+1.5\%($^{40}$K) & CNO+2\%($^{40}$K) \\
 \hline
 0.1 & 2997.64 & 5254.39 & 6334.59 & 7414.79 \\
 0.2 & 2736.43 & 4363.73 & 5142.65 & 5921.57 \\
 0.3 & 2447.10 & 3597.83 & 4148.63 & 4699.42 \\
 0.4 & 2129.82 & 2924.45 & 3304.78 & 3685.11 \\
 0.5 & 1787.78 & 2319.67 & 2574.24 & 2828.80 \\
 0.6 & 1439.26 & 1781.15 & 1944.77 & 2108.38 \\
 0.7 & 1100.80 & 1308.16 & 1407.39 & 1506.62 \\
 0.8 & 787.671 & 901.87 & 956.52 & 1011.16 \\
 0.9 & 526.847 & 579.58 & 604.81 & 630.04 \\
 1.0 & 355.736 & 371.46 & 378.98 & 386.5 \\
 1.1 & 249.9 & 249.9 & 249.9 & 249.9 \\
 1.2 & 156.71 & 156.71 & 156.71 & 156.71 \\
 1.3 & 81.17 & 81.17 & 81.17 & 81.17 \\
 1.4 & 26.75 & 26.75 & 26.75 & 26.75 \\
 1.5 & 0.303 & 0.303 & 0.303 & 0.303 \\
 \hline
\end{tabular}
\vspace{10mm}
\end{table}

\section{Reasons to use the Hydridic Earth model.}

\vspace{0.5cm}
\hspace{0.5cm} The widespread belief in the fairness of Silicate Earth model and belief in the validity of the results of work \cite{davis} that the heat flux from the Earth interior is equal to 47$\pm$2 TW do not allow to include in analysis a new source of single events in Borexino - the $^{40}$K geo-antineutrinos interaction.

The idea that the Hydridic Earth model is not correct is also widely accepted. One of the objections is that the entire Earth could have melt due to radiogenic heat and spent most part of its life in this state if the Earth contains potassium more than 1\% of the Earth weight. The current heat flux from the Earth's interior can be more than 200 TW in the frame of Hydridic Earth model which contradicts to the result of the work \cite{davis}.  
		
We have learned how to respond to the objections to Hydridic Earth model. The answers to the main objections are published in \cite{bezr2, bezr2018, barab2019, bezr22018}. Here we note that the entire Earth could not have melt because the Hydridic Earth model contains a subsurface cooling mechanism that is activated when the subsurface is heated enough to decompose the metal hydrides. Therefore, in the Hydridic Earth model, the subsurface temperature oscillates \cite{Larin}. 

It is also noted in \cite{bezr2, bezr2018, barab2019, bezr22018}  that thermal conductivity is not the main mechanism of heat transfer in the Earth, but protons and hydrogen-containing gases carry out the heat away. In these works the experimental evidences are provided that the heat flux from Earth interior can reach the several hundreds TW. These are the heating of the oceans, the temperature profile of ultra-deep wells and non-direct evidence – Moon heat flux from interior. 
		
Moreover, the effect was found that is predicted by the Hydridic Earth model and not predicted by the Silicate Earth model.
The Hydridic Earth model predicts that the Earth's crust is positively charged and contains a large amount of positive charge in the form of protons and positive ions of various hydrogen-containing gases. We tested the validity of this prediction experimentally by detecting an excess of the concentration of positive air-ions in underground rooms over the concentration of negative air-ions. In order to make this prediction, a new model of terrestrial electricity was developed based on the Hydridic Earth model \cite{bezr2019}. The model named "Hydridic model of terrestrial electricity".  This model explains the origin of the atmospheric electric field and all the observed effects of atmospheric electricity in a single way. The model also explains the origin of telluric currents. In \cite{bezr22019}, the Hydridic model of terrestrial electricity was successfully used in the analysis of the reaction of telluric currents to earthquakes.
 
 We consider Hydridic Earth model as an adequate tool for the study of the Earth. 

\vspace{0.5cm}

\section{Conclusion.}
\begin{enumerate}
    \item 
We propose to include the scattering of $^{40}$K geo-antineutrinos by electrons in the analysis of single event energy spectrum of Borexino detector. 
\item
The Hydridic Earth model predicts the potassium concentrations up to 4\% of the Earth mass. 
\item 
It is expected that the total interaction rate of the solar CNO neutrino and $^{40}$K geo-antineutrino will be obtained in the range of $6 \div 9$ events per day in 100 tons of scintillator.
\item	
We consider Hydridic Earth model as an adequate tool for the study of the Earth.  
\end{enumerate}

\section{Acknowledgment}
The authors express their gratitude to M. D. Skorokhvatov for the opportunity to discuss this note at a seminar at the NRC "Kurchatov Institute".

\end{document}